\documentclass[twocolumn]{aastex701}
\usepackage{amsmath}

\usepackage{fontspec}
\usepackage{xeCJK}
\usepackage{subcaption}

\shorttitle{Core shifts of BL Lacertae}
\shortauthors{Liu, Yang, Cheng et al.}

\begin{document}
\title{Detections of nearly bias-free core shifts with $5$--$30\,\mathrm{\mu as}$ precisions at 8--43 GHz in BL Lacertae}

\author[orcid=0000-0003-2778-002X,gname=Niu,sname=Liu]{Niu Liu (刘牛)}
\affiliation{School of Astronomy and Space Science, Nanjing University, Nanjing 210023, China}
\affiliation{Key Laboratory of Modern Astronomy and Astrophysics(Ministry of Education), Nanjing University, Nanjing 210023, China}
\email[show]{niu.liu@nju.edu.cn}

\author[orcid=0000-0002-2322-5232,gname=Jun, sname=Yang]{Jun Yang (杨军)} 
\affiliation{Department of Space, Earth and Environment, Chalmers University of Technology, Onsala Space Observatory, SE-43992 Onsala, Sweden}
\email[show]{jun.yang@chalmers.se}

\author[orcid=0000-0003-4407-9868,gname='Xiao-Peng',sname=Cheng]{Xiaopeng Cheng (程晓朋)}
\affiliation{Korea Astronomy and Space Science Institute, 776 Daedeok-daero, Yuseong-gu, Daejeon 34055, Republic of Korea}
\email[show]{xcheng@kasi.re.kr}

\author[orcid=0009-0000-9427-4608,gname=Ai-Ling,sname=Zeng]{Ai-Ling Zeng (曾艾玲)}
\affiliation{Instituto de Astrof\'{i}sica de Andaluc\'{i}a-CSIC, Glorieta de la Astronom\'{i}a s/n, E-18008 Granada, Spain}
\email{azeng@iaa.es}

\author[0000-0002-5519-0628]{Wen Chen (陈文)}
\affiliation{Yunnan Observatories, Chinese Academy of Sciences, Yangfangwang 396th, Kunming, China}
\email{chenwen@ynao.ac.cn}

\author[orcid=0000-0002-4439-5580,sname=Xiaolong,gname=Yang]{Xiao-Long Yang (杨小龙)}
\affiliation{Shanghai Astronomical Observatory, Chinese Academy of Sciences, Shanghai 200030, China}
\email{yangxl@shao.ac.cn}

\author[orcid=0000-0002-1992-5260,sname=Xiaoyu,gname=Hong]{Xiaoyu Hong (洪晓瑜)}
\affiliation{Shanghai Astronomical Observatory, Chinese Academy of Sciences, Shanghai 200030, China}
\affiliation{University of Chinese Academy of Sciences, Beijing 100049，China}
\email{xhong@shao.ac.cn}

\author[orcid=0009-0007-0631-291X,sname=Zhang,gname=Xia-Xuan]{Xia-Xuan Zhang (张夏轩)}
\affiliation{School of Astronomy and Space Science, Nanjing University, Nanjing 210023, China}
\affiliation{Key Laboratory of Modern Astronomy and Astrophysics(Ministry of Education), Nanjing University, Nanjing 210023, China}
\email{xiaxuanzhang@smail.nju.edu.cn}

\author[orcid=0000-0002-6637-9258,sname=Liu,gname=Jia-Cheng]{Jia-Cheng Liu (刘佳成)}
\affiliation{School of Astronomy and Space Science, Nanjing University, Nanjing 210023, China}
\affiliation{Key Laboratory of Modern Astronomy and Astrophysics(Ministry of Education), Nanjing University, Nanjing 210023, China}
\email{jcliu@nju.edu.cn}

\author[sname=Zhu,gname=Zi]{Zi Zhu (朱紫)}
\affiliation{School of Astronomy and Space Science, Nanjing University, Nanjing 210023, China}
\affiliation{Key Laboratory of Modern Astronomy and Astrophysics(Ministry of Education), Nanjing University, Nanjing 210023, China}
\affiliation{University of Chinese Academy of Sciences, Nanjing 211135, China}
\email{zhuzi@nju.edu.cn}


\begin{abstract}
When a radio jet is partially optically thick in the launching region, its apparent compact core may display frequency-dependent positional shifts. 
High-precision astrometric measurements of core shifts enable astronomers to pinpoint the jet's origin and place tight constraints on the magnetic field. 
BL Lacertae, the archetypal BL Lac object, hosts a highly variable and well-collimated jet. 
To independently constrain its innermost core shifts, we conducted very long baseline interferometric (VLBI) observations at 8.4, 12.4, 15.2, 23.6, and 43.2~GHz.
By exploiting a nearby (13.3 arcmin) steep-spectrum calibrator (NVSS~J220340+420839) through inverse phase-referencing VLBI astrometry, we detect nearly unbiased two-dimensional core shift measurements with state-of-the-art precisions of 5--30\,$\mu$as, which are significant at $>3\sigma$ confidence. 
The core shift between 8.4 and 43.2~GHz reaches 250\,$\mu$as.
The apparent core shifts scale with frequency as $\nu^{-1/k_r}$, implying the existence of an optically thick region in the upstream of jet. 
The derived core-shift index, $k_r\!=\!1.18^{+0.59}_{-0.34}$, is consistent, within uncertainties, with the canonical $k_r\!=\!1$ expected under energy equipartition between the jet particle and magnetic field energy densities, while allowing for modest deviations given that BL~Lacertae was captured in a flaring state.
\end{abstract}



\section{Introduction} \label{sec:intro}

Blazars are the most highly variable class of radio-loud active galactic nuclei, whose emission is dominated by relativistic jets oriented at small angles to the line of sight and strongly affected by Doppler boosting. 
Relativistic jets are thought to be launched through the interaction between a supermassive black hole (SMBH) and parsec-scale magnetic fields in the surrounding accretion flow \citep{2008Natur.452..966M}. 
Magnetic field is expected to play a key role in governing jet propagation, particularly within the inner jet acceleration and collimation region.  
Under some astrophysical jet models, the magnetic field near SMBHs can be properly probed with high-resolution very long baseline interferometry (VLBI) observations.  

On the VLBI map, the radio core is typically identified as the location where the optical depth due to synchrotron self-absorption approaches unity \citep{1979ApJ...232...34B}. 
Its apparent position therefore depends on observing frequencies, with lower-frequency cores appearing farther downstream than those observed at higher frequencies, following $r\!\propto\!\nu^{-1/k_r}$. 
This frequency-dependent displacement of the radio core, known as the core shift, provides an effective probe of physical conditions in the innermost jet \citep{1981ApJ...243..700K,1998A&A...330...79L,2009MNRAS.400...26O}. 
Previous core-shift measurements have often yielded power-law indices $k_r$ (also known as core-shift index) close to unity \citep{2009MNRAS.400...26O,2011A&A...532A..38S,2011Natur.477..185H}, broadly consistent with a freely expanding conical jet under near-equipartition between particle and magnetic energy densities \citep{1998A&A...330...79L}. 
Recent studies have found that core-shift variability is common in active galactic nuclei and may be associated with nuclear flares that inject denser plasma into the jet \citep{2019MNRAS.485.1822P,2023A&A...672A.130C}.
If the radio core includes a strong standing shock, the core positions at high frequencies might converge toward the shock location, resulting in little or no measurable core shifts \citep{Marscher2010}. 
Consequently, the core shift effect could help search for standing shocks in the inner jet region.

BL Lacertae is one of the best BL Lac objects for studying jet astrophysics close to SMBHs thanks to its proximity \citep[$z\!=\!0.0655$;][]{2022ApJS..261....2K} and brightness. 
High-angular-resolution VLBI observations reveal a bright core and a one-sided jet extending southward, with two stationary features near the core \citep{2008Natur.452..966M,2016ApJ...817...96G,2021A&A...649A.153C} when there is no giant flares.
The jet structure of BL Lacertae is, however, known to be highly variable with time, particularly during episodes of strong flaring activity \citep{2022Natur.609..265J}.
Using a two-dimensional cross-correlation alignment based on the optically thin jet emission in multi-frequency VLBI images, \citet{2009MNRAS.400...26O} measured significant core shifts (40--460 $\mu$as) between 4.6 and 43.1 GHz, consistent with expectations from the standard conical jet model. 
In addition, source-frequency phase-referencing VLBI observations reported a much smaller core shift of only $\sim20\,\mu$as between 22 and 43 GHz during a long-lived giant flare, possibly resulting from a recollimation shock in the inner region \citep{2017ApJ...834..177D}. 
Recently, we have identified a compact, steep-spectrum source at an angular separation of only $\sim 13.3^{\prime}$ from BL Lacerate in two pilot experiments including a Very Long Baseline Array (VLBA) experiment (project code: TC031; PI: Jun Yang) at 22~GHz and 43~GHz, and an European VLBI Network (EVN) experiment (project code: RSY08; PI: Jun Yang) at 5~GHz. 
This enables us to gain more accurate differential calibration of atmospheric and instrumental phase errors \citep{2020A&ARv..28....6R}, and provides independent, nearly unbiased measurements of core shifts in BL~Lacertae.

This paper is organized as follows. 
Section~\ref{sec:obs} describes the VLBI observations and data reduction. 
Section~\ref{sec:results} presents the multifrequency differential astrometry results for BL~Lacertae.
We compare our results with previous studies to constrain the nature of the jet magnetic field in Section~\ref{sec:discussion}. 
Finally, we summarize our conclusions in Section~\ref{sec:conclusion}. 
For a flat universe with $\Omega_\mathrm{ m}\!=\!0.315$, $\Omega_{\Lambda}\!=\!0.811$, and $H_0\!=\!67.4\,\mathrm{~km}\,\mathrm{~s}^{-1}\,\mathrm{Mpc}^{-1}$ \citep{2020A&A...641A...6P}, the redshift of BL Lacertae corresponds to a luminosity distance $D_L\!=\!305.59\,\mathrm{Mpc}$ and 1 mas corresponds to 1.39~parsec (pc).

\section{Observations and Data Reduction} \label{sec:obs}

We observed BL~Lacertae and the reference source NVSS~J220340+420839 (hereafter J2203+4208) with the VLBA (project code: BY196; PI: Jun~Yang) on 2024 July 27 in a phase-referencing observation mode at five frequencies: 8.4~GHz, 12.4~GHz, 15.2~GHz, 23.6~GHz, and 43.2~GHz. 
All VLBA antennas participated except Los Alamos. 
We followed a standard US National Radio Astronomy Observatory (NRAO) Astronomical Image Processing System (\textsc{aips}, version 31DEC25) procedure to calibrate visility data at all bands, using BL~Lacertae as fringe finder and bandpass calibrator. 
After preliminary calibration, BL~Lacertae was imaged in \textsc{Difmap} (version 2.5q). 
The source models (CLEAN components) determined from the BL~Lacertae images were exported to \textsc{AIPS} and used as the input model to derive self-calibration solutions (phase and amplitude), which were applied to both BL~Lacertae and J2203+4208.
The details about observational strategy, experiment setup and calibration procedure are summarized in Appendix~\ref{sec:obs_setup}.

The calibrated visibilities of J2203+4208 were then split and imaged in \textsc{Difmap} with natural weighting.
A single circular Gaussian component was fitted to the target using \textsc{modelfit}, yielding the integrated flux density and the component position relative to the phase center (Appendix~\ref{sec:modelfit}).
BL~Lacertae shows a compact, one-sided core-jet morphology oriented roughly north–south with core dominated (Fig.~\ref{fig:bllac_multifre_pos}). 
The core position of BL~Lacertae is defined as the location of the peak on the maps, for which non-zero offsets relative to the image origin (correlator phase center) were accounted for by model-fitting with a single elliptical Gaussian component to the core region using \textsc{jmfit} in \textsc{AIPS} (Appendix~\ref{sec:modelfit}). 
Subtracting this BL~Lacertae core positional offset from position offsets of J2203+4208 aligns cores of BL~Lacertae across frequencies onto a common local reference frame and provides core shift measurements.

\section{Multifrequency differential astrometry on BL Lacertae} \label{sec:results}

\subsection{Radio Spectrum of J2203+4208} \label{sec:j2203-spectra}

The reference source J2203+4208 was firmly detected at all five frequencies.
At frequencies of 23.6~GHz and below, the signal-to-noise ratios (SNRs) on the dirty maps exceeded 13, while the SNR was approximately 5.5 at 43.2~GHz.
The frequency dependence of the integrated flux densities obtained from the model fitting is shown in Appendix~\ref{sec:modelfit}.
A standard non-thermal power-law model was used to fit the spectrum of J2203+4208, in which the flux density $S_\nu$ at frequency $\nu$ is expressed as
%
\begin{equation} \label{eq:j2203-spectrum}
    S_\nu (\nu) = S_0 \cdot \nu^{\alpha},
\end{equation}
where $S_0$ is the amplitude of the synchrotron spectrum and $\alpha$ is the spectral index.
The fitting yielded $\alpha\!=\!-1.27\,\pm\,0.02$, indicating a steep spectrum of J2203+4208.
Therefore, J2203+4208 is optically thin and hence not subject to any detectable core shift, making it a reliable reference source for measuring the core shift of nearby targets.

\subsection{Nearly unbiased core shift measurements} \label{sec:bllac-core-shift}

\begin{figure*}[ht]
    \centering
    \includegraphics[width=\textwidth]{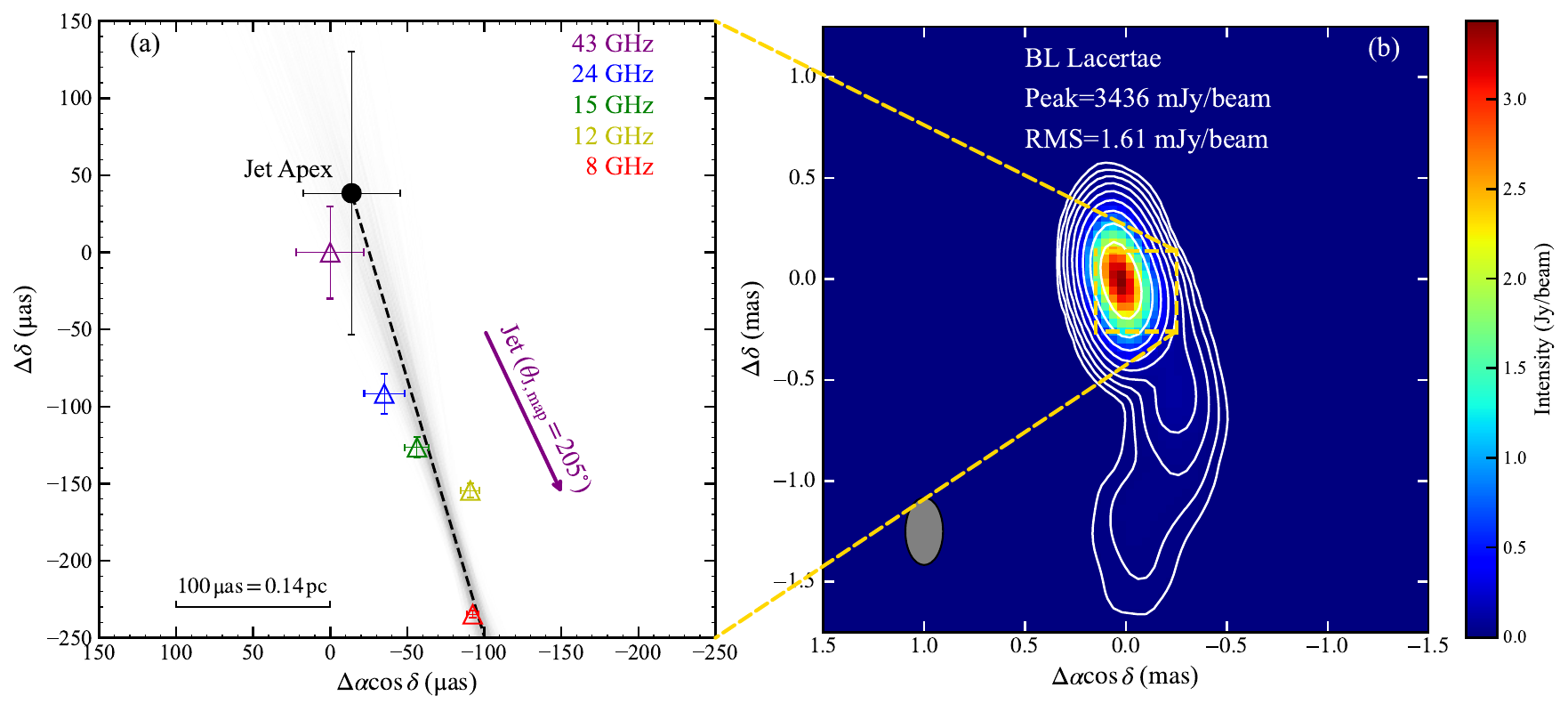} 
    \includegraphics[width=0.48\textwidth, height=0.345\textwidth]{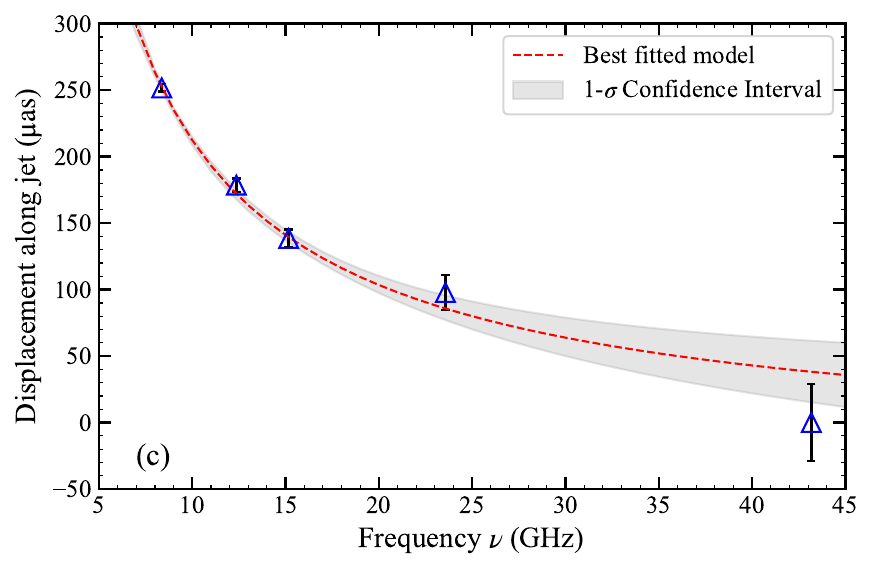} 
    \includegraphics[width=0.51\textwidth]{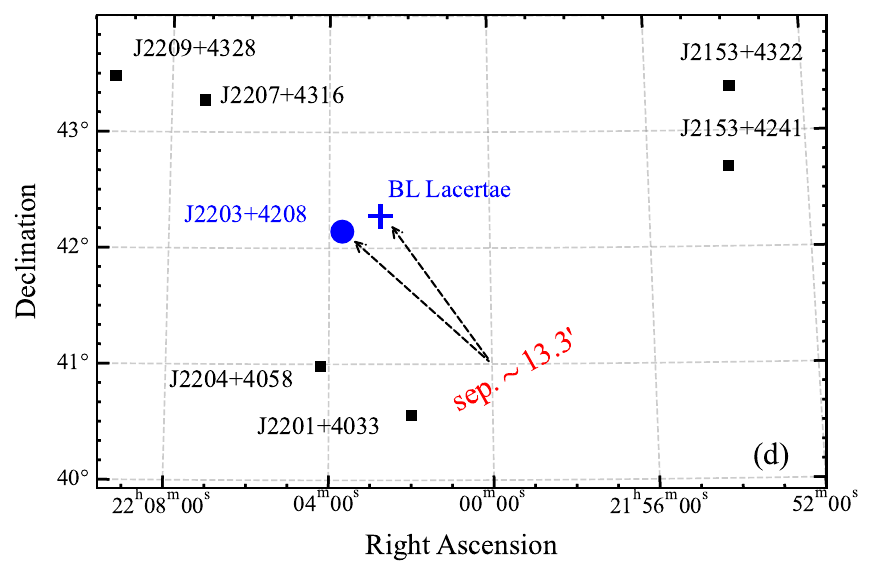} \\
    \caption{\label{fig:bllac_multifre_pos} %
    Core shift measurements of BL~Lacertae.
    (a) Two-dimensional core shifts at 8.4, 12.4, 15.2, and 23.6~GHz relative to 43.2~GHz.
    The dashed line denotes the best-fit core-shift model, and the gray region shows the scatter from 1000 randomly sampled MCMC realizations.
    (b) 43~GHz VLBA image.
    (c) Projected core displacements along the jet direction, adopting a position angle of $197^{\circ}$.
    The red dashed line shows the best-fit model prediction, with the gray region indicating the 1$\sigma$ confidence interval.
    (d) Calibrator field around BL~Lacertae.
    Black squares mark six sources from the Radio Fundamental Catalog \citep[rfc\_2025c;][]{2025ApJS..276...38P}, and the blue circle indicates the reference calibrator used in this work.
  }
\end{figure*}

The relative positional offsets of J2203+4208 were determined with a precision of $\mathrm{\leq 7\,\mu as}$ at frequencies lower than 15.2~GHz, $\mathrm{13\,\mu as}$ at 23.6~GHz, and approximately $\mathrm{30\,\mu as}$ at 43.2~GHz.
During our observations, the mean zenith angle of J2203+4208 and BL~Lacertae is $27^{\circ}$ at 43.2~GHz and ranges from $40^{\circ}$ to $50^{\circ}$ at other frequencies. 
Taking into account a residual tropospheric zenith excess path length of 3~cm as adopted in \cite{2020A&ARv..28....6R}, the systematic errors due to extrapolation of phase referencing solutions are all below $\mathrm{5\,\mu as}$ and thus can be reasonably omitted in the following analyzes.
Assuming that the position of J2203+4208 does not change with frequency, we aligned positions of the innermost core of BL~Lacertae onto the 43~GHz map and determined the core shift of BL~Lacertae, as presented in Table~\ref{tab:bllac_core_shift}.
We also measured the position angle of jet direction (counted from the north to east) using the 43~GHz calibrated visibilities of BL~Lacertae by \textsc{modelfit}.
The position angle of the first elliptical-Gaussian component was adopted, which gives $\theta_\mathrm{J,map}\!=\!205.36\,\pm\,0.01^{\circ}$.
As shown in Fig.~\ref{fig:bllac_multifre_pos}, the core positions of BL~Lacertae at five frequencies roughly align, which is nearly parallel to the direction of jet.

\begin{table}[hpbt]
    \centering
    \caption{\label{tab:bllac_core_shift}
         Core shift measurements for BL~Lacertae.
        }
    \begin{tabular}{ccccc}
        \hline\hline
        Freq. &$\Delta\alpha^*$ &$\Delta \delta$ &Shift &PA \\
        (GHz) &($\mathrm{\mu as}$)  &($\mathrm{\mu as}$) &($\mathrm{\mu as}$)  &(${}^{\circ}$) \\
        \hline
        \multicolumn{5}{c}{43.2~GHz core as origin} \\
        8.4  &$-92\pm4$ &$-235\pm2$ &$253\pm29$ &$202\pm5$ \\
        12.4  &$-91\pm6$ &$-155\pm5$ &$181\pm28$ &$210\pm8$ \\
        15.2  &$-56\pm8$ &$-126\pm7$ &$141\pm29$ &$204\pm11$ \\
        23.6  &$-35\pm13$ &$-92\pm13$ &$102\pm31$ &$201\pm17$ \\
        43.2  &$0\pm22$ &$0\pm30$ &- &- \\
        \multicolumn{5}{c}{23.6~GHz core as origin} \\
        8.4  &$-57\pm4$ &$-143\pm2$ &$155\pm13$ &$202\pm5$ \\
        12.4  &$-56\pm6$ &$-63\pm5$ &$85\pm14$ &$222\pm10$ \\
        15.2  &$-21\pm8$ &$-35\pm7$ &$44\pm14$ &$211\pm22$ \\
        \multicolumn{5}{c}{15.2~GHz core as origin} \\
        8.4  &$-36\pm4$ &$-108\pm2$ &$114\pm7$ &$198\pm4$ \\
        12.4  &$-35\pm6$ &$-28\pm5$ &$45\pm9$ &$231\pm12$ \\
        \multicolumn{5}{c}{12.4~GHz core as origin} \\
        8.4  &$-2\pm4$ &$-80\pm2$ &$80\pm5$ &$181\pm5$ \\
    \hline
    \end{tabular}
    \tablecomments{
    The formal uncertainties of $\Delta\alpha^*\!=\!\Delta\alpha\cos\delta$ and $\Delta \delta$ represent the positional uncertainties of core at each individual frequency for BL~Lacertae, while the computation of formal uncertainties in shift and position angle (PA) have taken the positional uncertainties at both two frequencies into consideration.
    }
\end{table}

We adopted a two-dimensional model to describe the core shift at the observing frequency $\nu$ as
\begin{equation} \label{eq:core-shift-ra-dec}
    \begin{aligned} 
        \Delta\alpha\cos\delta &= x_0 + r_0 \cdot \nu^{-\kappa} \sin \theta_{\rm J} ,\\
        \Delta \delta &= y_0 + r_0 \cdot \nu^{-\kappa} \cos \theta_{\rm J},
    \end{aligned}
\end{equation}
where $r_0$ and $\kappa\!=\!1/k_r$ are core shift parameters, and $\theta_{\rm J}$ is the core shift direction counted from the north to east.
The coordinate ($x_0$, $y_0$) of the jet apex can also be directly inferred from the multifrequency relative astrometric measurements without an assumed jet direction.
We fitted the observed astrometric position offsets with the physical core-shit model given in Eq.~(\ref{eq:core-shift-ra-dec}) and determined the posterior distributions of the parameters
$(r_{0},\,k_r,\,x_{0},\,y_{0},\,\theta_{\rm J})$ with a Markov Chain Monte Carlo (MCMC) sampler.
This approach accounts for the non--linear nature of the model and strong covariances among parameters (Appendix~\ref{sec:mcmc_fit}).
The fit returned
\begin{equation} \label{eq:fit_results}
    \begin{aligned} 
        r_0 &=1726^{+1123}_{-425}\,\mathrm{\mu as}, \quad k_r = 1.18^{+0.59}_{-0.34},\\
        x_0 &=-14^{+34}_{-21}\,\mathrm{\mu as}, \quad y_0 =38^{+100}_{-60}\,\mathrm{\mu as}, \\
        \theta_{\rm J} &=197\pm 3^{\circ}.
    \end{aligned}
\end{equation}
This estimate of direction of the core shift aligns well with the average value of $14^{\circ}\,\pm\,10^{\circ}$ reported by \citet{2009MNRAS.400...26O}.
The apparent difference reflects the fact the position angles are defined modulo $180^{\circ}$; our value of $197^{\circ}$ is equivalent to $17^{\circ}$ and is consistent within the formal uncertainties.

\section{Discussion} \label{sec:discussion}

The projected core shifts of $30\,\mu\mathrm{as}$ to $180\,\mu\mathrm{as}$ (typical uncertainty of 20--$40\,\mu\mathrm{as}$) relative to 43.1~GHz determined in \citet{2009MNRAS.400...26O} are slightly smaller than our measurements, but remain consistent within the formal uncertainties.
\citet{2012A&A...545A.113P} measure core shifts in 2006 with position angles of $\sim 180^{\circ}$.
While the core shift of $\sim70\,\mu\mathrm{as}$ to $\sim130\,\mu\mathrm{as}$ between 8.4~GHz and 15.4~GHz encompasses our value ($114\pm7\,\mu\mathrm{as}$),
their shift between 12.1~GHz and 15.4~GHz (124--$160\,\mu\mathrm{as}$) is significantly larger than our measurement of $45\pm9\,\mu\mathrm{as}$.
\citet{2017ApJ...834..177D} report a shift of approximately $20\,\mu\mathrm{as}$ between 22~GHz and 43~GHz, which is significantly smaller than the prediction of the core shift model.
We measure a shift of approximately $100\,\mu\mathrm{as}$ between 23.6~GHz and 43.2~GHz with a $>3\sigma$ confidence.
We note that our observations coincided with a strong flaring state of BL~Lacertae.
An further inspection of the MOJAVE (15~GHz) \footnote{\url{https://www.cv.nrao.edu/MOJAVE/sourcepages/2200+420.shtml}} and BEAM-ME (43~GHz)\footnote{\url{https://www.bu.edu/blazars/VLBA_GLAST/bllac.html}} light curves reveals that previous results correspond to different activity states compared to the strong flare seen in our data, likely explaining the discrepancies.
Similar variability-dependent core shift behavior has also been observed in other sources \citep[e.g.,][]{2019MNRAS.485.1822P}.
Since the frequency dependence of the measured core shifts in our data is well described by the core shift model, we conclude that the 43~GHz core observed in our experiment corresponds to an optically thick component.

For an equipartition state between the jet particle and magnetic field energy densities, the core shift index is expected to be $k_r\!=\!1$.
Previous reported values are generally consistent with this assumption, including $k_r=0.99\pm0.07$ from \citet{2009MNRAS.400...26O} and $k_r=1.06\pm0.22$ in \citet{2017MNRAS.469..813A} derived from radio light curves at five frequencies between 4.8~GHz and 36.8~GHz spanning more than 40~yr.
The derived value of the core shift index of 1.18 remains higher, but is still consistent with this assumption, considering the formal uncertainty.
The excess may be caused by the flare, as observed in 3C 345.3 \citep{2023A&A...672A.130C}.

We calculate the magnetic field strength from the core shift measurements following \citet{2009MNRAS.400...26O}, using the revised formula from \citet{2023A&A...672A.130C}.
The jet kinematic parameters are taken from \citet{2005AJ....130.1418J}, including the jet viewing angle of $7.7^{\circ}$, the Doppler factor of 7.2, and the half-opening angle of the jet of $1.9^{\circ}$.
Aadopting $k_r=1$ yields a core magnetic field strength $B_{\mathrm{core}}(\nu)$ ranging from $0.06\,\pm\,0.01\,\mathrm{G}$ at 8.4~GHz to $0.26\,\pm\,0.05\,\mathrm{G}$ at 43.2~GHz.
This corresponds to a magnetic field strength at 1\,pc of $B_1\!=\!0.134\,\pm\,0.006\,\mathrm{G}$, roughly consistent with $B_1\!=\!0.139\,\mathrm{G}$ in \citet{2009MNRAS.400...26O} while greater than 0.09~G given in \citet{2014Natur.510..126Z}.
Alternatively, using $k_r=1.18$ yields slightly lower high-frequency magnetic fields ($0.23\,\pm\,0.03\,\mathrm{G}$ at 43.2~GHz), resulting in a higher magnetic field strength at 1\,pc of $B_1\!=\!0.201\pm0.010\,\mathrm{G}$.
Based on these results, we predict an 86~GHz core magnetic field strength of $0.32\,\mathrm{G}$ (for $k_r=1.18$) to $0.42\,\mathrm{G}$ (for $k_r=1$).

While the core-shift measurements broadly follow the jet direction, we notice a small deviation from a strictly linear, one-dimensional trend in Fig.~\ref{fig:bllac_multifre_pos}.
Given the high phase-referencing precision of our measurements, this non-linear alignment may reflect the intrinsic structure (e.g., jet bending) or the frequency-dependent opacity effects coupled with variations in the jet direction (i.e., jet precession) of BL~Lacertae. 
Further multi-frequency observations would be required to confirm its astrophysical origin.

We also note that the calibrator, J2203+4208, exhibits a tentative jet-like extension toward the north (Appendix~\ref{sec:modelfit}), which could potentially introduce systematic astrometric shifts due to core-jet blending. 
To quantify this effect, we performed a trial re-analysis by introducing an additional component to represent the possible jet emission.
The resulting change in the relative core position of BL Lacertae is found to be less than $30\,\mu\mathrm{as}$ in both right ascension and declination.
This structural blending is most pronounced at 8.4~GHz, whereas no such feature is apparent at 23.6 and 43.2~GHz.
Re-deriving the core-shift parameters via MCMC fitting with these updated positions yielded results fully consistent with our primary analysis within the formal uncertainties. 
Therefore, we conclude that the impact of core-jet blending in the calibrator is negligible and does not affect our conclusions. 
Further dedicated VLBI observations at lower frequencies would help to confirm the nature of this tentative feature of the calibrator.

\section{Conclusion} \label{sec:conclusion}

We present, for the first time, two-dimensional core shift measurements at the precisions of $5\,\mathrm{\mu as}$ to $30\,\mathrm{\mu as}$ for BL~Lacertae between 8.4 and 43.2~GHz, obtained using the inverse phase-referencing technique.
This straightforward approach enables us to derive nearly bias-free core shift measurements, particularly between 23.6 and 43.2~GHz, owing to the small angular separation of approximately $13^{\prime}$ between BL~Lacertae and the reference steep-spectrum source (approximately an order of magnitude smaller than in previous studies).
These results are entirely independent of earlier work and provide new constraints on the physical conditions of the jet in BL~Lacertae during its flaring state.

\begin{acknowledgments}

NL and JCL were supported by the National Natural Science Foundation of China (NSFC) under grant Nos.~12573070 and 12373074.
W.~Chen was supported by the National Natural Science Foundation of China (NSFC) under grant No. 12573073, Yunnan Fundamental Research Projects (grant No.~202401AT070144) and Yunnan Foreign Talent Introduction Program (grant No.~202505AO120021).
XC was supported by the Brain Pool Program through the National Research Foundation of Korea (NRF) funded by the Ministry of Science and ICT (RS-2024-00407499).
This research was supported by the National Research Foundation of Korea(NRF) funded by the Korea government(KASA, Korea AeroSpace Administration) (grant numbers RS-2024-00509838 and R25TA0065942000). 
This work used data from the MOJAVE database that is maintained by the MOJAVE team \citep{2018ApJS..234...12L} and the Radio Fundamental Catalog (rfc\_2025c) in our analysis with the associated dataset available at NASA \citep{https://doi.org/10.25966/dhrk-zh08}.
This work also uses VLBA data from the VLBA-BU Blazar Monitoring Program (BEAM-ME and VLBA-BU-BLAZAR; http://www.bu.edu/blazars/BEAM-ME.html), funded by NASA through the Fermi Guest Investigator Program. 
The VLBA is an instrument of the National Radio Astronomy Observatory. 
The National Radio Astronomy Observatory is a facility of the National Science Foundation operated by Associated Universities, Inc.
The European VLBI Network is a joint facility of independent European, African, Asian, and North American radio astronomy institutes. 
The scientific results from the data presented in this publication are derived from the following EVN project code(s): RSY08.
This work benefited from the Box, LaTeX, and Git services provided by the e-Science Center of the Collaborative Innovation Center of Advanced Microstructures, Nanjing University.
The code used to reproduce all results in this paper is publicly available at \url{https://git.nju.edu.cn/astrometry/bllac_core_shift}.

\end{acknowledgments}

\facilities{VLBA (NRAO)}

\software{\textsc{AIPS} \citep{2003ASSL..285..109G}, 
          \textsc{Difmap} \citep{1997ASPC..125...77S},
          \texttt{astropy} \citep{2013A&A...558A..33A,2018AJ....156..123A,2022ApJ...935..167A}, 
          \texttt{numpy} \citep{2011CSE....13b..22V},
          \texttt{SciPy} \citep{2020NatMe..17..261V},
          \texttt{emcee} \citep{2013PASP..125..306F},
          \texttt{matplotlib} \citep{2007CSE.....9...90H},
          Jupyter Notebook \citep{2016ppap.book...87K},
          }


\appendix

\section{Observation strategy and calibration process} \label{sec:obs_setup}

Thanks to the small angular separation of approximately $13.3^{\prime}$ between BL Lacertae and J2203+4208, we allocated 18~s for the switching between sources in each cycle.
The duration times were approximately 1~hr at 8.4~GHz, 1~hr at 12.4~GHz, 1~hr at 15.2~GHz, 2~hr at 23.6~GHz, and 6~hr at 43.2~GHz (11~hr total).
Table~\ref{tab:setup} summarizes the nodding cycle times and observational setup.

Data were recorded in dual polarization at 4096~Mb\,s$^{-1}$ with 2-bit quantization, using $4\times128$~MHz subbands per polarization.
Correlation was performed with the DiFX software correlator in Socorro, New Mexico, using 0.5~s integration and 0.5~MHz spectral channels.
The correlation position was set to $\mathrm{R.A.}\!=\!22^\mathrm{h}02^\mathrm{m}43^\mathrm{s}.2914$, $\mathrm{Dec.}\!=\!+42^{\circ}16^{\prime}39^{\prime\prime}.980$ for BL Lacertae and to $\mathrm{R.A.}\!=\!22^\mathrm{h}03^\mathrm{m}40^\mathrm{s}.6110$, $\mathrm{Dec.}\!=\!+42^{\circ}08^{\prime}39^{\prime\prime}.419$ for J2203+4208,
the latter been determined from a VLBA testing observation experiment (project code: TC031; PI: Jun Yang) at 23.6~GHz and 43.2~GHz and an EVN short-observation (project code: RSY08; PI: Jun Yang) at 4.9~GHz.

The calibration followed the standard \textsc{aips} procedures and was identical at all bands.
Ionospheric dispersive delays were corrected via global total electron content maps derived from Global Navigation Satellite System data.
The residual delays due to inaccurate a priori values of earth orientation parameters (EOPs) were corrected using the standard rapid EOP solutions from the International Earth Rotation and Reference Systems Service.
Phase terms from antenna parallactic angle variations were removed.
Instrumental single-band delays and phase offsets were determined from a dedicated two-minute scan of BL~Lacertae and applied to all scans. 
The bandpass responses were derived from the BL~Lacertae scans, averaging in time over the full track and in frequency within each 128\,MHz subband; one solution per intermediate frequency (IF) and polarization was applied. 
A~priori amplitude calibration used the measured system temperatures and antenna gain curves recorded at each station during the run. 
After data inspection and flagging, global fringe fitting was performed on BL~Lacertae with a 10\,s solution interval at 8.4--23.6\,GHz and 5\,s at 43.2\,GHz, using a point-source model and averaging across all subbands. 
Phase solutions from BL~Lacertae were then transferred to J2203+4208 via linear interpolation in time.

\begin{figure}[ht]
  \centering
  \includegraphics[width=\textwidth]{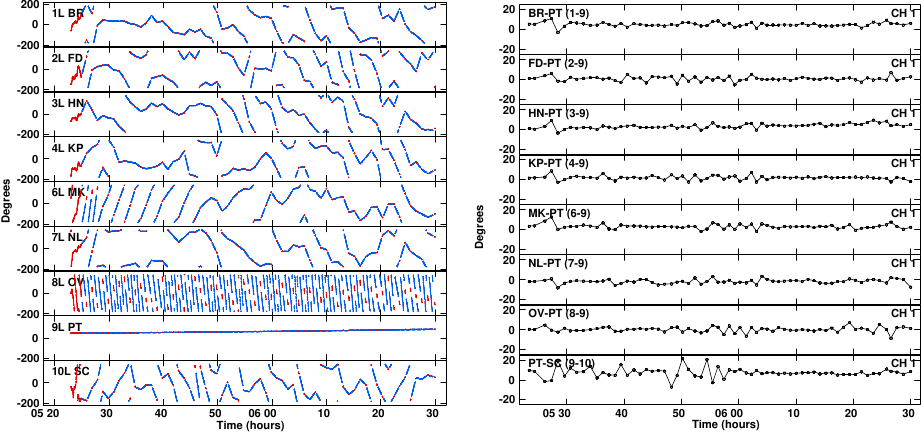} 
  \caption{\label{fig:43GHz_cl13} %
  Left: Final phase solutions at 43.2~GHz over the first observing hour during the first hour of observations.
  Red symbols denote the phase solutions of BL~Lacertae, and blue symbols show the interpolated phases transferred from BL~Lacertae to J2203+4208.
  Right: Residual phases for BL~Lacertae after application of the final phase solutions at 43.2~GHz during the same time interval.
  }
\end{figure}

After preliminary calibration, the visibility data of BL~Lacertae were exported, after flagging 32 edge channels (25\%), and then imaged in \textsc{Difmap}. 
The source model of BL~Lacertae were exported to \textsc{AIPS} and used as the input model to derive self-calibration solutions. 
These solutions were applied to both BL~Lacertae and J2203+4208.
The left panel of Fig.~\ref{fig:43GHz_cl13} presents the final phase solutions for all stations at 43.2~GHz during the first observing hour. 
The rapid phase variations of the VLBA-OV (Owens Valley) station are observed at all five frequencies, which are likely to be caused by the residual clock rate after correlation.
The right panel shows the residual phases for BL~Lacertae after applying the final phase solutions at 43.2~GHz over the same time interval, demonstrating small phase residuals.

\begin{table}[htbp]
    \centering
    \caption{
        Nodding cycle times, on-source scan lengths, and total on-source integration times for BL~Lacertae (BL\,Lac) and J2203+4208 (J2203).}
    \label{tab:setup}
    \begin{tabular}{cccccc}
        \hline\hline
        Freq. & Nodding &\multicolumn{2}{c}{On-source scan} &\multicolumn{2}{c}{Total on-source time} \\
              & Cycle      &BL\,Lac  &J2203 &BL\,Lac  &J2203 \\
        (GHz) & (min) & (s)   &(s)      & (min)   &(hr)\\
        \hline
        8.4  & 4 & 10 & 210 & 4.0   & 1.05 \\
        12.4 & 3 & 10 & 150 & 7.5   & 0.92 \\
        15.2 & 3 & 10 & 150 & 7.5   & 0.83 \\
        23.6 & 2 & 25 &  90 & 12.5  & 1.25 \\
        43.2 & 1 &  5 &  40 & 31.9  & 3.72 \\
    \hline
    \end{tabular}
\end{table}

\section{Model-fitting results} \label{sec:modelfit}

\begin{figure}[ht]
  \centering
  \includegraphics[width=\textwidth]{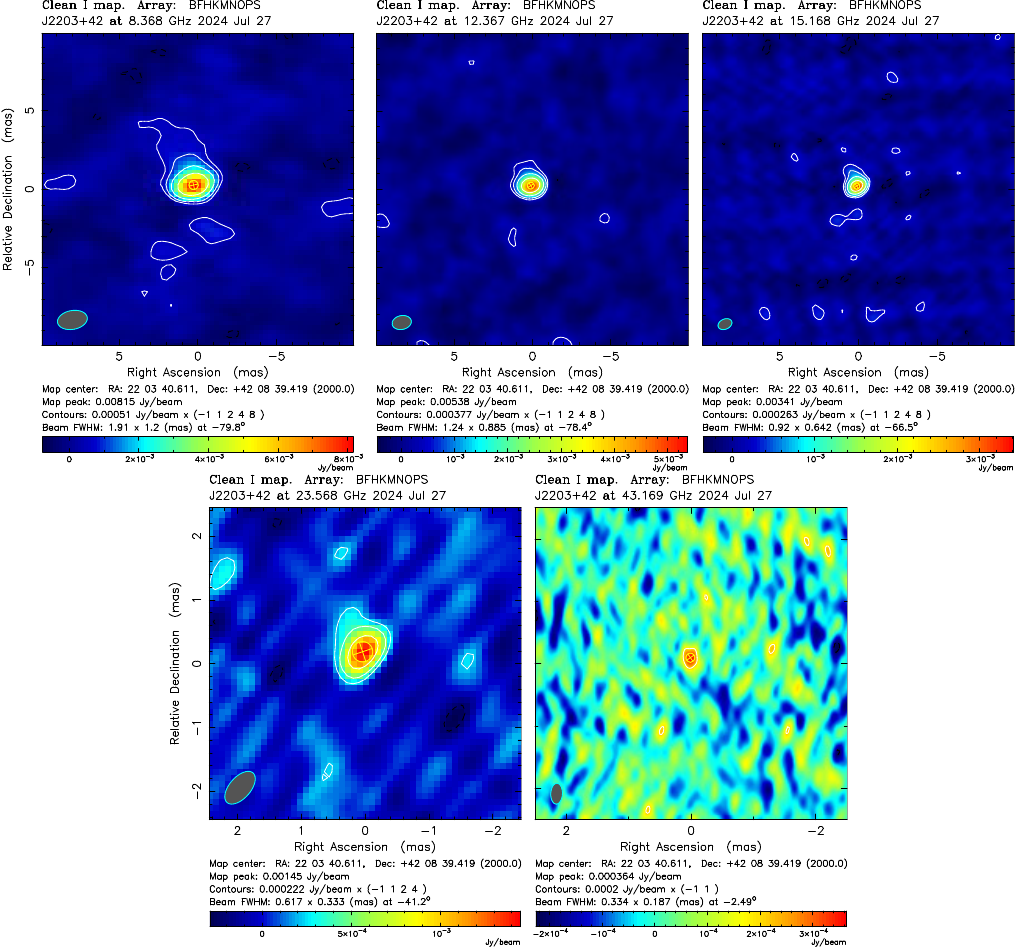} 
  \caption{\label{fig:image_j2203} %
  VLBA maps of J2203+4208 at 8.4~GHz, 12.4~GHz, 15.2~GHz, 23.6~GHz, and 43.2~GHz, respectively.
  The contours start from $\mathrm{3\sigma}$ off-source noise level.
  }
\end{figure}

\begin{table}[ht!]
    \centering
    \caption{\label{tab:j2203_modelfit}
        Model-fitting results of J2203+4208.}
    \begin{tabular}{cccccc}
        \hline\hline
        Freq. &$\Delta\alpha^*$ &$\Delta \delta$ &$S_{\rm int}$ &$S_{\rm peak}$ &FWHM\\
        (GHz) &($\mathrm{\mu as}$)  &($\mathrm{\mu as}$) &(mJy) &($\mathrm{mJy\,beam^{-1}}$) &(mas)\\
        \hline
        8.4  &$273\pm4$ &$267\pm2$ &$9.45\pm0.04$ &$8.15\pm0.17$ &0.576 \\
        12.4  &$110\pm6$ &$230\pm5$ &$6.38\pm0.08$ &$5.38\pm0.13$ &0.428 \\
        15.2  &$65\pm8$ &$208\pm6$ &$4.24\pm0.09$ &$3.41\pm0.09$ &0.362 \\
        23.6  &$41\pm13$ &$188\pm13$ &$2.27\pm0.13$ &$1.45\pm0.07$ &0.318 \\
        43.2  &$12\pm22$ &$98\pm30$ &$0.50\pm0.12$ &$0.36\pm0.07$ &0.142 \\
    \hline
    \end{tabular}
    \tablecomments{The last column is the full width at half maximum.}
\end{table}

Figure~\ref{fig:image_j2203} displays the CLEAN maps of J2203+4208 at all five frequencies. 
The peak SNRs in the CLEAN maps are approximately 47 at 8.4~GHz, 43 at 12.4~GHz, 39 at 15.2~GHz, 20 at 23.6~GHz, and 5.5 at 43.2~GHz.
A single circular-Gaussian model was adopted to fit the calibrated visibility data of J2203+4208 in order to determine its integrated flux densities and relative positions with respect to the reference position using the \textsc{modelfit} task.
The results of model-fitting for J2203+4208 are summarized in Table~\ref{tab:j2203_modelfit}.
We further fitted the total integrated flux densities of J2203+4208 with the model in Eq.~(\ref{eq:j2203-spectrum}). 
The non-linear least-squares optimization was performed using the \texttt{curve\_fit} routine in the \texttt{SciPy} module, adopting the Levenberg--Marquardt algorithm. 
The fit yields an estimate of $S_{0} = 140 \pm 6~\mathrm{mJy}$.

\begin{figure}[ht!]
  \centering
  \includegraphics[width=0.5\textwidth]{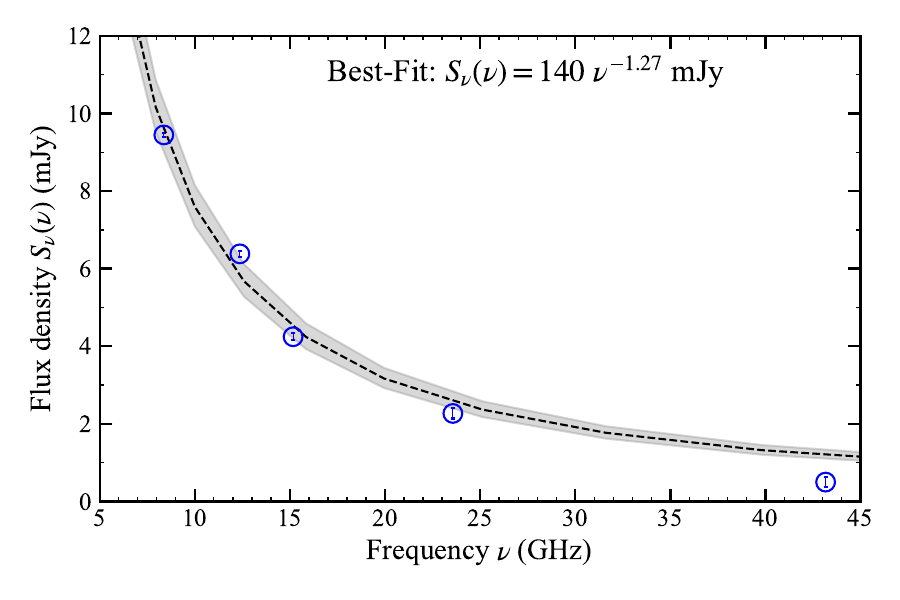} 
  \caption{\label{fig:j2203_spectrum} %
  Radio spectrum of the reference source J2203+4208. 
  The blue open circles denote the integrated flux densities obtained in this work, while the gray zone indicates the 1-$\sigma$ confidence interval.
  }
\end{figure}

\begin{figure}[ht]
  \centering
  \includegraphics[width=\textwidth]{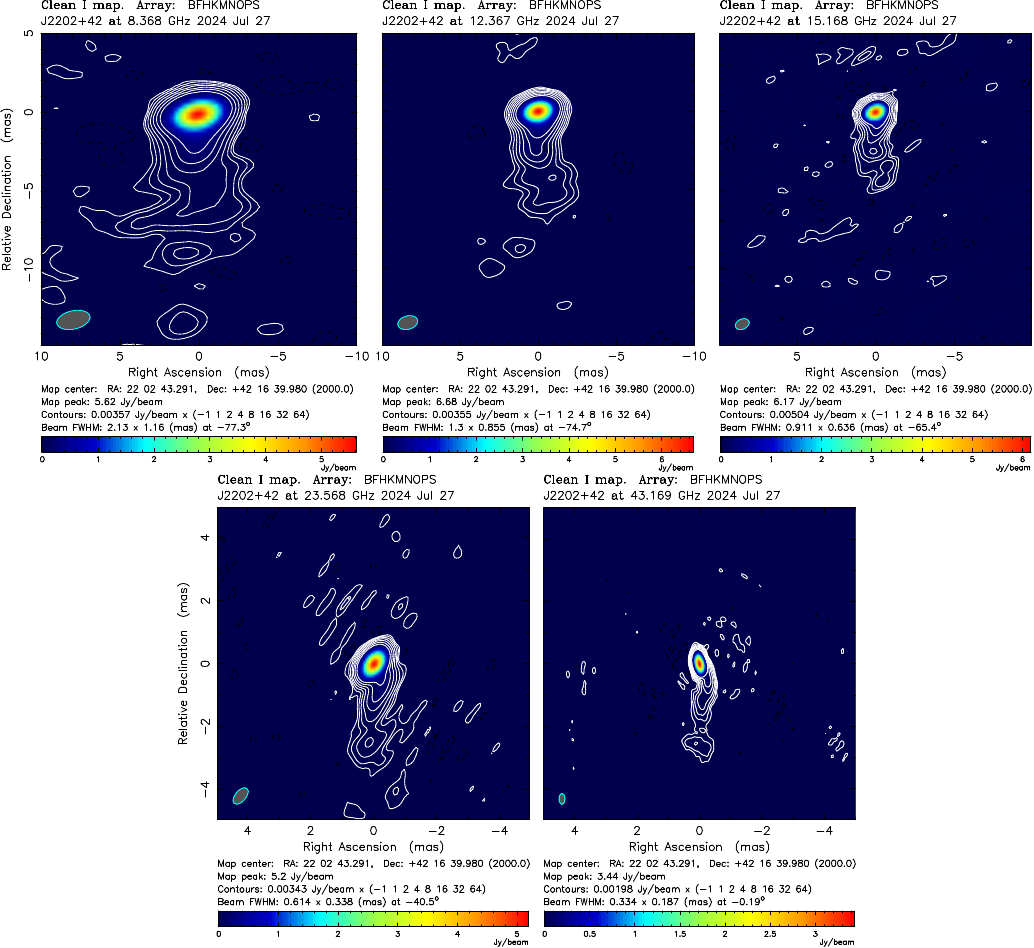} 
  \caption{\label{fig:image_bllac} %
  VLBA maps of BL~Lacertae at 8.4~GHz, 12.4~GHz, 15.2~GHz, 23.6~GHz, and 43.2~GHz, respectively.
  The contours start from $\mathrm{3\sigma}$ off-source noise level.
  }
\end{figure}

\begin{table}[ht!]
    \centering
    \caption{\label{tab:bllac_modelfit}
        Model-fitting results of BL~Lacertae. 
        }
    \begin{tabular}{cccccccc}
        \hline\hline
        Freq. &$\Delta\alpha^*$ &$\Delta \delta$ &$S^{\rm core}_{\rm int}$ &$S_{\rm peak}$ &$a$  &$b$  &$\phi$\\
        (GHz) &($\mathrm{\mu as}$)  &($\mathrm{\mu as}$) &(Jy) &($\mathrm{Jy\,beam^{-1}}$) &(mas) &(mas) &(${}^{\circ}$)\\
        \hline
        8.4  &$174$ &$-55$ &$6.164\pm0.002$ &$5.680\pm0.001$ &0.456  &0.301  &187\\
        12.4  &$12$ &$-12$ &$7.098\pm0.002$ &$6.703\pm0.001$ &0.300  &0.062  &205\\
        15.2  &$3$ &$-6$ &$6.777\pm0.004$ &$6.256\pm0.002$   &0.277  &0.138  &222\\
        23.6  &$-1$ &$9$ &$6.546\pm0.002$ &$5.209\pm0.001$   &0.265  &0.076  &207\\
        43.2  &$5$ &$10$ &$5.615\pm0.001$ &$3.450\pm0.001$   &0.346  &0.052  &205\\
    \hline
    \end{tabular}
    \tablecomments{The formal uncertainties on the positional offsets are typically $\lesssim0.1\,\mathrm{\mu as}$. 
    $a$ and $b$ are the major and minor axes FWHM of the Gaussian component and $\phi$ is the position angle of the major axis of the component, measured North through East.}
\end{table}

    Figure~\ref{fig:image_bllac} yields the VLBA maps of BL~Lacertae at five frequencies.
    We used the \textsc{jmfit} task in \textsc{aips} to estimate the relative positional offset of peaks on the maps with respect to the correlation phase center coordinate for BL~Lacertae.
    A model of single elliptical Gaussian component was used.
    Table~\ref{tab:bllac_modelfit} reports the model-fitting results. 
    The formal uncertainties on the positional offsets are typically $\lesssim0.1\,\mathrm{\mu as}$, which are not statistically meaningful and thus not listed in the table.

\section{Core Shift Parameter Estimation} \label{sec:mcmc_fit}

\begin{figure}[ht!]
    \centering
    \includegraphics[width=\linewidth]{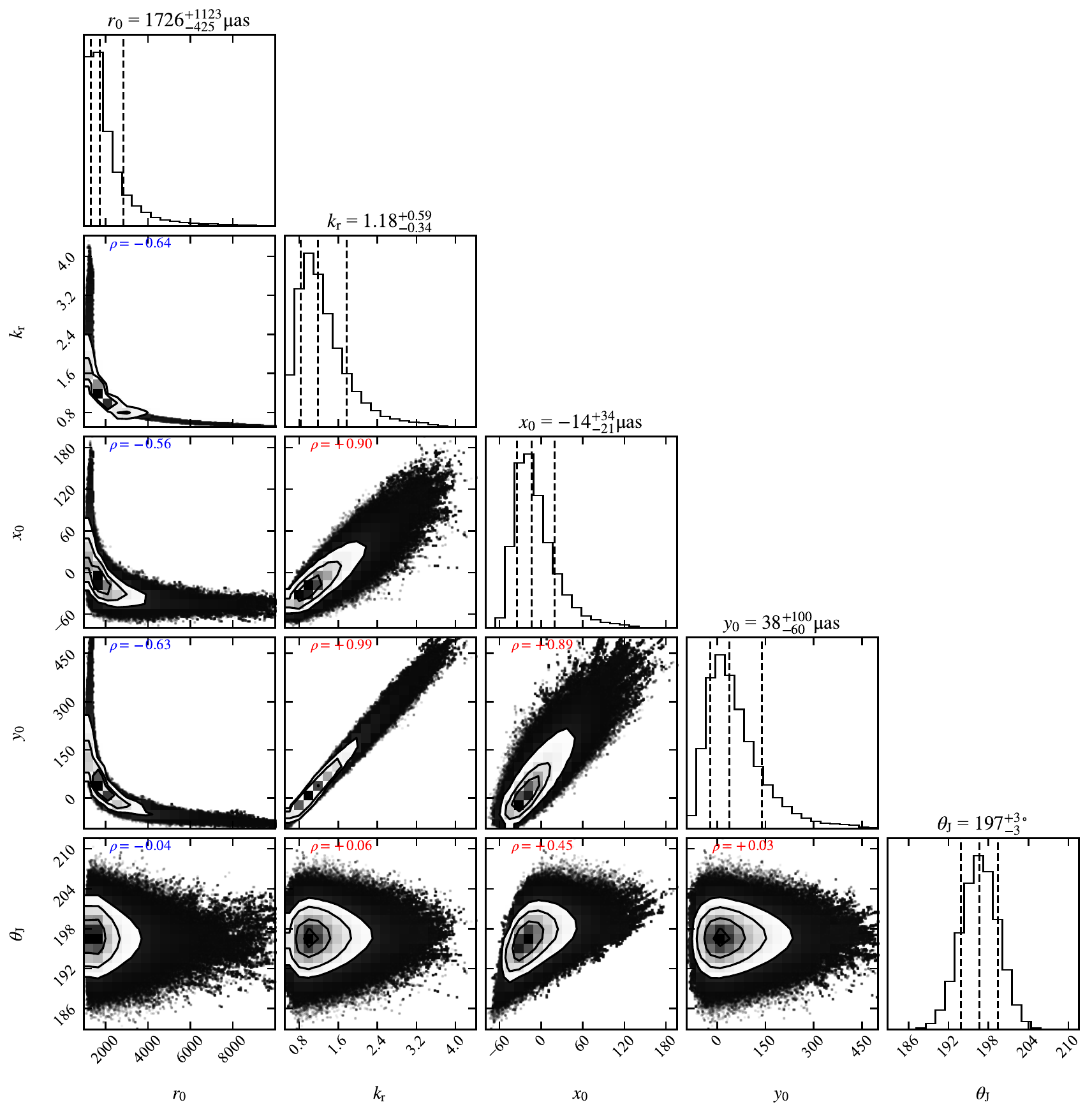}
    \caption{Posterior probability distributions of the core shift parameters derived from the MCMC analysis. 
    Pearson correlation coefficients between parameters are indicated, with red and blue denoting positive and negative correlations, respectively.}
    \label{fig:mcmc_post_dist}
\end{figure}

    In the MCMC fitting, we used $N_{\mathrm{walk}}=32$ walkers and ran each chain for $N_{\mathrm{step}}=2\times10^{5}$ steps. 
    The first $N_{\mathrm{burn}}=2\times10^{4}$ steps were discarded as burn-in, and the remaining samples were used to construct the posterior probability distributions. 
    For all parameters, the total chain length exceeds $50\,\tau_{\mathrm{int}}$, where $\tau_{\mathrm{int}}$ is the integrated autocorrelation time, indicating reliable convergence of the MCMC chains.
    We adopted the following priors: $0<r_{0}<10~\mathrm{mas}$, $0.1<k_r<10$, uniform priors on $(x_{0},\,y_{0})$ over $\pm5$~mas, and a uniform prior on the jet position angle of jet $\theta_{\rm J}\in[0,2\pi)$.
    Figure~\ref{fig:mcmc_post_dist} presents the posterior probability distributions of all model parameters, revealing strong correlations (absolute Pearson correlation coefficients exceeding 0.4) among most parameters, except for the jet position angle.

\bibliographystyle{aasjournalv7}
\bibliography{references}{}



\end{document}